# The Neural Basis of Groove Sensations: Implications for Music-Based Interventions and Dance Therapy in Parkinson's Disease

Chen-Gia Tsai[1,2*] and Chia-Wei Li[3]

[1] Graduate Institute of Musicology, National Taiwan University, Taipei, Taiwan
[2] Graduate Institute of Brain and Mind Sciences, National Taiwan University, Taipei, Taiwan
[3] Department of Radiology, Wan Fang Hospital, Taipei Medical University, Taipei, Taiwan
*Contact: tsaichengia@ntu.edu.tw

## Abstract

Groove sensations arise from rhythmic structures that evoke an urge to move in response to music. While syncopation has been extensively studied in groove perception, the neural mechanisms underlying low-frequency groove remain underexplored. This fMRI study examines the role of the mirror neuron system and associated brain regions in processing low-frequency groove. Region-of-interest analysis revealed that amplifying drum and bass components in K-pop songs significantly increased activity in the right posterior inferior frontal gyrus, right inferior/superior parietal lobules, left dorsolateral prefrontal cortex, and bilateral posterior middle/inferior temporal gyrus. These findings suggest that low-frequency grooves engage sensorimotor, executive, and rhythm semantics networks, reinforcing their role in action-related processing. Building on these insights, we propose an enhanced rhythmic auditory stimulation paradigm for Parkinson's disease, incorporating amplified low-frequency rhythmic cues to improve gait synchronization.

## 1. Introduction

The term "groove" in music refers to both the rhythmic structures that evoke a compelling urge to move and the subjective experience of wanting to move. At least two key musical features contribute to groove sensations. First, rhythmic complexity, particularly deviations from an expected metrical framework (e.g., syncopation), has been shown to induce bodily movement (Vuust *et al.*, 2018) and is associated with activation of the dorsal auditory pathway (Morillon *et al.*, 2019; Zalta *et al.*, 2024). Second, the prominence of low-frequency sounds, typically produced by drum and bass components, strongly enhances movement motivation. Electronic Dance Music, for example, is characterized by powerful bass drops that often elicit pronounced physical responses (Dayal & Ferrigno, 2012; Solberg & Jensenius, 2017).

While the Action Simulation for Auditory Prediction hypothesis and the Gradual Audiomotor Evolution hypothesis provide explanations for understanding how syncopation induces groove (Proksch *et al.*, 2020), the theoretical framework for low-frequency groove remains relatively underdeveloped. This gap in understanding has contributed to the scarcity of neuroimaging studies on this type of groove sensation. Our recent findings suggest that the mirror neuron system (MNS) serves as a key neural substrate underlying low-frequency groove sensation (Li & Tsai, 2024). The MNS is a network of brain regions, including the posterior inferior frontal gyrus (pIFG), the adjacent ventral premotor cortex (vPMC), and the inferior parietal lobule (IPL), which are involved in action perception and motor resonance by mapping observed movements onto corresponding motor representations (Rizzolatti & Craighero, 2004). Notably, this system is also engaged in mapping heard movements onto motor representations (Le Bel *et al.*, 2009), suggesting that the MNS may facilitate the coupling between rhythmic auditory input and motor system activation, thereby contributing to groove sensations.

This study aims to elucidate the neural mechanisms underlying low-frequency groove sensation and seeks to translate these findings to enhance music-based interventions and dance therapy for Parkinson's Disease (PD). Specifically, we hypothesize that incorporating low-frequency rhythmic emphasis into Rhythmic Auditory Stimulation (RAS) protocols will provide stronger external cues for gait regulation, compensating for striatal-cortical loop deficits in PD.

## 2. Methods

To investigate the neural substrates underlying the processing of low-frequency rhythmic prominence in music, we performed an fMRI study with K-pop song stimuli. Twenty-five participants were subjected to fMRI scans while listening to musical stimuli. A detailed description of the experimental methods is available in our previously published paper (Li & Tsai, 2024); here, we provide a brief overview of the stimuli and introduce a novel analysis of the fMRI data.

Three distinct versions of each 15-second excerpt were generated: (1) Whole Music – the original recording, (2) Drum & Bass – a version containing only the drum and bass tracks, and (3) Vocals & Others – a version in which the drum and bass tracks were removed, leaving only the vocals and other instrumental components. The 18 pop song excerpts used in this study can be found at https://youtu.be/EXqvdFbprH0 (accessed on 5 May 2024). While listening to Whole Music primarily allows for the appreciation of the song's emotional content and musical aesthetics, particularly the vocals (although none of the participants understood the lyrics), listening to Drum & Bass emphasizes the perception of low-frequency grooves.

In our previous study (Li & Tsai, 2024), we employed a one-way ANOVA to identify statistically significant differences in neural activation across the three conditions. Regions that exhibited significant effects in the ANOVA were subsequently used as an inclusive mask to constrain further analyses. To delineate specific pairwise differences between conditions, post-hoc paired *t*-tests were conducted within these masked regions. However, in the present study, we adopt an *a priori*-defined region-of-interest (ROI) approach to mitigate the risk of false negatives that may arise when restricting subsequent analyses to ANOVA-derived inclusive masks. This approach ensures that our investigation is guided by established neuroanatomical and functional hypotheses, thereby enhancing the sensitivity of our analyses to detect relevant neural effects.

The following ROIs were selected based on their well-documented involvement in action planning, sensorimotor integration, and motor representation: right pIFG/vPMC, bilateral IPL, left dorsolateral prefrontal cortex (DLPFC), right superior parietal lobule (SPL), and bilateral posterior middle/inferior temporal gyrus (MTG/ITG). Each of these ROIs plays a distinct yet interrelated role in action-related neural processing. Right pIFG/vPMC, right IPL, and left IPL are critically involved in action perception (Spunt & Lieberman, 2012). The left DLPFC is hypothesized to function as a higher-order controller that exerts top-down modulation over the MNS by selecting and integrating elementary motor representations (Buccino *et al.*, 2004). The right SPL plays a pivotal role in sensorimotor integration, visuospatial processing, and attentional control, all of which are essential for supporting MNS-mediated action understanding and execution (Wang *et al.*, 2015a). The bilateral posterior MTG/ITG facilitate the recognition and interpretation of learned movement patterns, as evidenced by their increased activation when professional dancers observe familiar dance styles, likely reflecting their role in mapping motion onto stored visuomotor representations (Calvo-Merino *et al.*, 2005).

We conducted paired *t*-tests within *a priori* defined ROIs (4 mm-radius spheres centered on coordinates from prior literature; Table 1) to compare neural responses between the Whole Music and Drum & Bass conditions. Beta values were extracted from each ROI, and statistical significance was determined with a Bonferroni correction for multiple comparisons.

**Table 1.** Regions of Interest (ROIs) used in the present study, including their MNI coordinates, associated prior studies, and experimental contrasts from which peak coordinates were derived.

| Brain Region | X | Y | Z | Study | Contrast |
|---|---|---|---|---|---|
| Right pIFG | 57 | 13 | 13 | Montgomery *et al.* (2007) | Viewing, imitating and producing communicative hand gestures > rest |
| Right IPL | 58 | -35 | 41 | Montgomery *et al.* (2007) | Viewing, imitating and producing communicative hand gestures > rest |
| Left IPL | -60 | -46 | 40 | Becchio *et al.* (2012) | Observation of grasping actions performed with social versus individual goals |
| Left DLPFC | -46 | 32 | 36 | Peak coordinates derived from a meta-analysis of "planning" using Neurosynth | N/A |
| Right SPL | 29 | -46 | 59 | Bray *et al.* (2015) | Spatial working memory manipulation |
| Right posterior MTG/ITG | 54 | -54 | -8 | Molnar-Szakacs *et al.* (2006) | Stacked subassembly (high action semantic processing) > seriated subassembly (low action semantic processing) |
| Left posterior MTG/ITG | -52 | -58 | -6 | Yang and Shu (2016) | Non-literal action language of fictive motion, motion-related semantic representation |

## 3. Results

ROI analysis revealed significantly greater activation in the right pIFG/vPMC, right IPL, right SPL, left DLPFC, and bilateral posterior MTG/ITG during the Drum & Bass condition compared to the Whole Music condition. No significant difference in activation was observed in the left IPL (Figure 1).

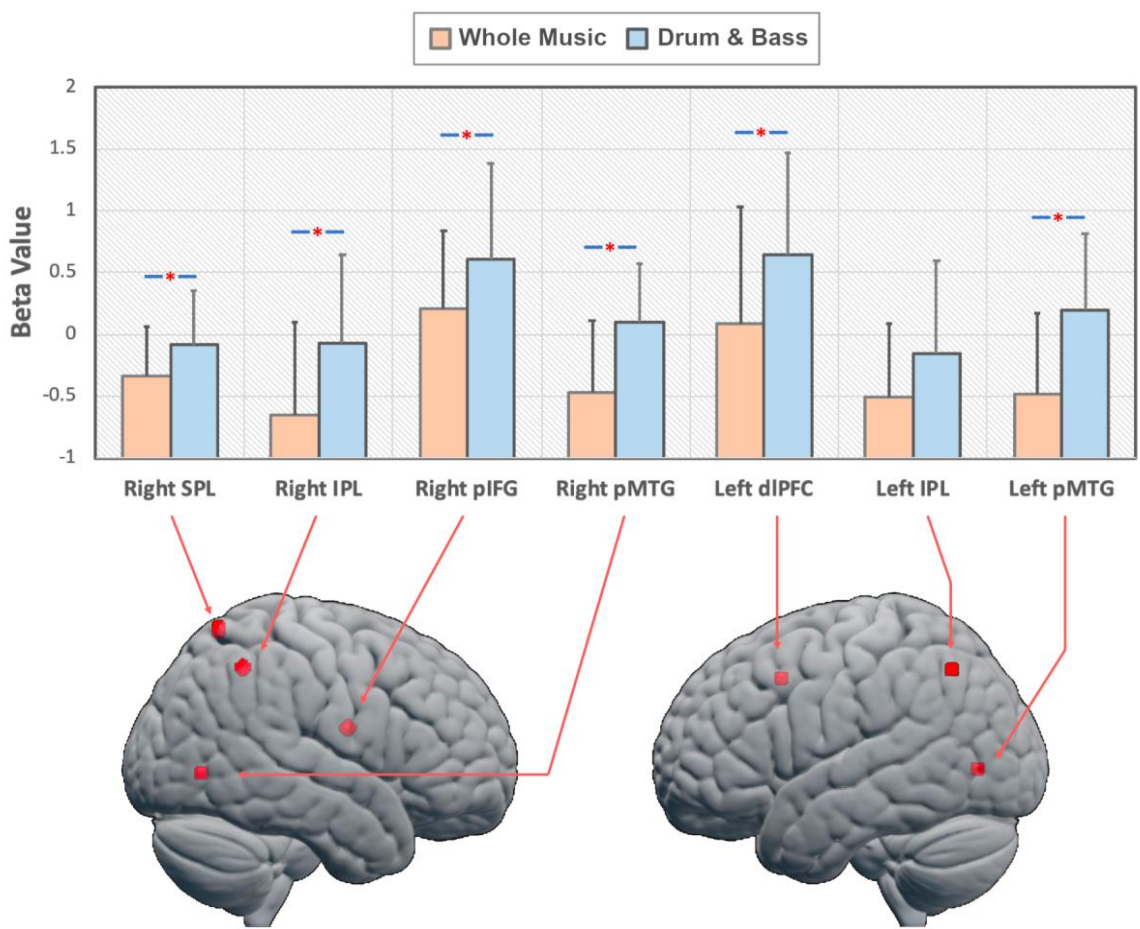

**Figure 1.** Results of the ROI analysis. Error bars represent the standard error, and asterisks indicate statistically significant differences.

## 4. Discussion

### 4.1. Frontoparietal MNS

The MNS is thought to facilitate social communication and musical engagement by mapping observed or heard actions onto an individual's own motor representations (Molnar-Szakacs & Overy, 2006). Prior research has shown that the pIFG/vPMC, key components of the MNS, link specific sounds to corresponding motor actions (Bangert *et al.*, 2006; Lahav *et al.*, 2007; Chen *et al.*, 2008). Our findings reveal that enhancing low-frequency groove in pop music elicits increased activity in the right pIFG/vPMC and IPL. This supports our hypothesis that the MNS plays a crucial role in transforming low-frequency auditory input into motor commands. The parietal motor cortex may be affected by cognitive processes in the prefrontal cortex when auditory stimuli correspond to motor actions (Li *et al.*, 2023). Our finding that the right IPL and SPL showed greater activity for Drum & Bass than Whole Music suggests that these parietal regions in right hemisphere may interact with the pIFG and DLPFC to process the kinesthetic imagery induced by low-frequency grooves.

### 4.2. Right IPL/SPL and Sensorimotor Integration

Both the IPL and SPL are integral to kinesthetic imagery, which involves mentally simulating movement sensations (Lebon et al., 2018). Additionally, the right IPL and SPL play a crucial role in sensorimotor integration and spatial processing (Wolpert et al., 1998; Kravitz et al., 2011; Lester & Dassonville, 2014), which are essential for constructing a mental representation of movement in space during motor imagery. Auditory cues, such as bass lines, may provide implicit spatial information that contributes to this representation.

In this study, fluctuations in the bass line might be perceived by the participant's parietal lobe as kinesthetic cues, representing bodily movement in space. Prior research has demonstrated a strong association between pitch height and vertical spatial representation (Rusconi et al., 2005). Notably, this pitch-height mapping appears to be language-independent, suggesting that such cross-modal associations reflect a universal perceptual tendency rather than a linguistic construct (Parkinson et al., 2012). We speculate that the right IPL and SPL may play a role in linking vertical fluctuations in the bass line to kinesthetic representations.

This effect may be subtle when the bass line is weak, but as it becomes stronger and more resonant, listeners experience an intensified sense of bodily drive. Such a response could stem from interactions between the vestibular and auditory systems. The human vestibular system is particularly sensitive to low-frequency vibrations and infrasound, with the otolithic organs playing a key role in modulating vestibular reflexes in response to such stimuli (Todd et al., 2009). Moreover, studies suggest that the vestibular and auditory systems integrate low-frequency information, influencing self-motion perception (Shayman et al., 2020). The right IPL and SPL may function as integration hubs, combining spatial information, vestibular signals, motor cortex activity, and proprioceptive feedback to facilitate bodily motion perception in response to low-frequency grooves. Future research should investigate how vertical fluctuations in bass lines influence groove sensations.

## 4.3. Left DLPFC and Motor Planning

The left DLPFC exhibited significantly greater activation in response to Drum & Bass compared to Whole Music. Previous work indicates that this region subserves a wide spectrum of executive functions, many of which directly involve movement planning. Rozzi and Fogassi (2017) highlight that the lateral prefrontal cortex (including the DLPFC) encodes both sensory information and behavioral meaning. Wang *et al.* (2015b) suggested that the DLPFC appears to bridge sensory inputs and prospective actions over time, playing a key role in the temporal organization of goal-directed behavior. Buccino *et al.* (2004) proposed that the DLPFC is critical for selecting and integrating elementary motor representations within the MNS. Additionally, Lage et al. (2015) emphasized the DLPFC's involvement in motor performance and memory processing, especially under conditions demanding variability, continuous adjustment, and strategic planning. In the present study, the repetitive yet variable drum patterns in Drum & Bass likely triggered dynamic motor simulation, engaging the DLPFC not only to select and combine motor representations but also to constantly adapt motor plans in response to auditory cues.

## 4.4. Bilateral Posterior MTG/ITG and Rhythm Semantics

We observed more pronounced activity in the bilateral posterior MTG/ITG for Drum & Bass compared to Whole Music. These regions are integral components of the ventral visual stream, which is critical for processing complex visual information (Kravitz *et al.*, 2013). The left posterior MTG/ITG, in particular, has been implicated in action semantics, playing a key role in the observation of expressive whole-body movement sequences (Tipper *et al.*, 2015) and the comprehension of figurative motion sentences (e.g., the road runs along the coast) (Romero Lauro *et al.*, 2013; Yang & Shu, 2016). Building on these findings, we propose that the bilateral posterior MTG/ITG plays a central role in the high-level pattern recognition and elaboration of low-frequency grooves. If groove perception relied solely on auditory cortical analysis and MNS-based motor resonance, one might be able to detect the rhythmic structure and exhibit a general tendency toward movement, yet without necessarily identifying and interpreting the groove as a specific dance or movement style. Consequently, this would limit the capacity to generate contextually appropriate motor responses. The posterior MTG/ITG bridges this gap in "rhythm semantics", extracting not only relevant motor concepts but also associated visual memory representations. Thus, the processing of low-frequency grooves necessitates the integrated functioning of the posterior MTG/ITG, the MNS, and auditory cortical regions, collectively forming a perception-action loop that enables the recognition, interpretation, and imagery of groove-related movements.

## 4.5. Rethinking the MNS, Social Coordination, and Groove Sensations

The MNS maps perceived actions onto an individual's own motor representations, thereby facilitating movement imitation and learning. However, the concept of action intention understanding within MNS theory has been challenged (Hickok, 2009), leading many researchers to favor the term Action Observation System (AOS) over MNS. AOS focuses on the neural processing of action observation, emphasizing operationally defined, measurable behaviors and mechanisms, thereby reducing controversy and improving empirical validation. While we acknowledge the strong link between MNS and motor learning and control, we also believe its broader functional roles should not be overlooked. Just as feathers originally evolved for insulation but later enabled flight, the MNS may have initially evolved for motor learning and control but later adapted to serve a role in social communication through biological and cultural evolution. Music has long served as a social communication mechanism, with the MNS playing a central role in this process. In the rock song "We Will Rock You", for instance, people stomp in synchrony with the bass drum, reinforcing social bonding. Such phenomena suggest that the MNS, rather than the AOS, better explains the neural basis of movement imitation induced by auditory cues, as AOS primarily concerns visual action observation.

It is interesting to note that other species also exhibit movement-based social coordination. For instance, many flocking bird species engage in pre-flight behaviors such as increased vocalizations, wing-flapping, and short hops, collectively functioning as a threshold-based (or quorum-like) mechanism that triggers a coordinated takeoff once enough individuals exhibit readiness (Sirot, 2006). Similarly, dolphins preparing to hunt may use pre-hunt vocalizations and synchronized movements to organize their group actions (Lopez Marulanda *et al.*, 2021). Drawing a parallel to human culture, low-frequency groove may serve a similar social signaling function, akin to a silent invitation among dancers—"Do you feel our rhythm? Join us!"—and it may explain why low-frequency drumming has historically facilitated collective entrainment in rituals, dances, and social gatherings, as well as in modern music festivals, concerts, and clubs. In such contexts, synchronous movement enhances the predictability of others' actions and fosters empathy, ultimately strengthening social trust and group cohesion. Importantly, resonating with another's movement does not necessarily imply deep social cognition or action understanding at a higher level. Rather, we emphasize that in musical and dance contexts, the ability to predict and feel others' movements plays a fundamental role in social-emotional bonding, a process that is primarily mediated by the MNS.

Another concept that warrants reconsideration is that groove sensations not only refer to the urge to move in response to music but also frequently encompass an associated sense of pleasure. A study on musical anhedonia challenges the assumption that these two components are inextricably linked. Specifically, Romkey *et al.* (2025) found that individuals with musical anhedonia, despite exhibiting a reduced capacity to experience pleasure from music, still demonstrated a desire to move that was comparable to controls. This finding suggests that the drive to move in response to rhythmic input may operate independently of hedonic reward mechanisms, implying that musical pleasure and the urge to move could represent partially dissociable psychological processes. Our research supports this dissociation, as we observed that different versions of K-pop groovy songs—Whole Music, Drum & Bass, and Vocals & Others—likely elicited distinct groove sensations, yet participants' preference ratings did not differ significantly across conditions (Li & Tsai, 2024).

# 5. Implications for PD and Rhythmic Auditory Stimulation

PD is a neurodegenerative disorder characterized by motor impairments, including bradykinesia, rigidity, tremor, and gait disturbances. These symptoms primarily result from the progressive loss of dopaminergic neurons in the basal ganglia, which disrupts internal timing mechanisms and motor control processes necessary for smooth, coordinated movement. Gait disturbances are among the most disabling motor symptoms of PD, significantly impacting mobility and quality of life. Some PD patients develop self-invented strategies, known as compensation strategies, to improve their gait (Tosserams *et al.*, 2023). Among these strategies, motor imagery and action observation involve activating the MNS, potentially facilitating cortically driven movement. Observing or mentally simulating another person's gait pattern may enhance motor planning and execution. Additionally, external cueing introduces goal-directed movement by providing rhythmic or spatial references, such as walking to the beat of music, which has been shown to improve gait initiation and step synchronization in PD patients.

Rhythmic Auditory Stimulation (RAS) has been proposed as an external cue to mitigate these deficits by providing a stable temporal framework for movement. By synchronizing their movements to regular auditory beats, patients can effectively bypass or enhance the weakened striatal-cortical loop (for a recent review, see Wang et al., 2022). RAS engages neural pathways implicated in motor control, including the PMC and the DLPFC. A previous study investigating the neural mechanisms of RAS found that, compared to healthy individuals, PD patients exhibit significantly increased functional connectivity between the auditory cortex and the executive control network when synchronizing their movements to a steady auditory cue (Braunlich *et al.*, 2019). The executive control network, which includes the DLPFC and the IPL, plays a key role in motor planning and coordination.

Conventional RAS protocols primarily rely on mid-to-high-frequency auditory cues to regulate gait. However, such stimuli may not fully leverage the potential benefits of low-frequency rhythmic elements, which are more closely aligned with natural biomechanical processes. For example, military marches and parade music prominently feature bass drums and low brass instruments to provide rhythmic cues that synchronize collective footsteps, reinforcing movement coordination through low-frequency auditory stimulation. Based on this principle, we propose an enhanced RAS paradigm that incorporates amplified low-frequency rhythmic components: **the bass drum simulates both footstep impact and body vibrations, while the bass line represents the sensation of foot-ground contact.** This approach may provide more intuitive and biomechanically relevant cues for gait regulation, particularly for individuals with movement disorders such as PD.

We propose that this low-frequency-enhanced RAS may offer distinct advantages over conventional RAS protocols for the following reasons:

➢ **Low-Frequency Grooves More Closely Resemble Natural Biomechanical Cues**

Natural footfalls generate low-frequency sounds and internal body vibrations that travel through the skeletal structure, including the feet, pelvis, and spine. Critically, the decay of these sounds and vibrations mirrors that of bass drum sounds. This similarity suggests that low-frequency auditory cues, compared to mid- or high-frequency cues, may more effectively entrain gait by resonating with the body's natural biomechanical rhythms.

➢ **Low-Frequency Grooves May Compensate for Impaired Plantar Sensory Feedback**
PD patients often experience reduced tactile sensitivity in the feet, impairing their ability to detect ground contact during gait. Additionally, their IPL exhibits diminished responsiveness to tactile stimuli (Huang *et al.*, 2025). Bass-line cues, which provide stronger sensory-motor integration signals, may help compensate for this deficit by enhancing IPL activation and reinforcing the perception of foot-ground contact.

➢ **Low-Frequency Grooves May Enhance Downbeat Perception**
Low-frequency grooves in pop music play a fundamental role in shaping meter perception. Across various music and movement traditions, including marching band music, ballet, and hip-hop dance, the moment of stepping or landing from a jump frequently aligns with the downbeat, the metrically strongest beat in a measure. This synchronization likely reflects the body's intrinsic need for stability upon impact, requiring the precise coordination of multiple sensory inputs to maintain balance. Tactile feedback from the feet, proprioceptive signals from the lower limbs, and vestibular cues from the saccule all converge to support postural adjustments during movement. Since landing demands precise multisensory integration, humans may naturally align this critical biomechanical event with the downbeat. By amplifying low-frequency rhythmic components in pop music, RAS may more effectively engage these sensorimotor mechanisms.

➢ **The Bass Line Resembles the Foot in Providing Support**
The bass line in music provides harmonic support, much like how the feet alternately bear weight during walking. Dawe *et al.* (1993) demonstrated that harmonic rhythm, or the rate of chord changes, influences meter perception, as significant chord transitions often occur on the downbeat of a measure. This aligns with the rhythmic structure of human gait, where shifts in body weight and foot placement typically correspond to the downbeat.

➢ **Enhanced Activation of the DLPFC and IPL May Facilitate Motor Control**
For PD patients experiencing difficulty initiating gait, low-frequency grooves may provide a form of "virtual" stepping feedback. That is, drum-and-bass cues might be misinterpreted by the brain as self-generated footfalls, creating a prediction error that compels the motor system to "correct" by initiating stepping movements. DLPFC and IPL may participate in processing these prediction errors (Hertrich *et al.*, 2021; Wang *et al.*, 2022). Once gait initiation begins, the low-frequency groove may engage the DLPFC and IPL to maintain step synchronization with auditory cues.

# 6. Conclusion and Perspectives

Neuroimaging evidence suggests that low-frequency groove sensation is mediated by the MNS, SPL, DLPFC, and posterior MTG/ITG. Enhancing low-frequency rhythmic elements in pop songs may increase activation in key motor-related brain regions that play a key role in rehabilitation for PD patients. Therefore, we propose an enhanced gait training approach, detailed in the Appendix.

Future research should compare how PD patients synchronize their stepping to engaging songs versus a monotonous metronome beat. These studies should investigate differences in neural activation patterns, stepping synchronization performance, and emotional responses to gain deeper insights into the potential benefits of music-based interventions. Additionally, individual differences in the effectiveness of this therapy should be explored, as factors such as musical preference, baseline motor function, and cognitive engagement may influence patient outcomes.

# Appendix: Enhanced RAS Protocol with Low-Frequency Groove for Gait Training

Objective: To improve gait synchronization and motor control in patients with movement disorders (e.g., Parkinson's Disease) by integrating rhythmic entrainment, action observation, motor imagery, and auditory-motor coupling.

## Step 1: Preparation

1. Select Music
   - Use familiar pop songs with a joyful melody to enhance engagement and motivation.
2. Adjust Tempo
   - Ensure the music tempo aligns with the patient's comfortable walking speed (cadence).
   - If needed, gradually adjust the tempo to facilitate smoother gait transitions.
3. Modify Drum and Bass Elements
   - Bass drum stroke should coincide with footstep impact.
   - Bass line should align with the sensation of foot-ground contact to reinforce proprioceptive awareness.
   - Amplify the drum and bass components.

## Step 2: Motor Preparation (Cognitive & Observational Training)

1. Action Observation & Motor Imagery
   - Patients watch a short video (e.g., https://rb.gy/2zdw51) of a person naturally walking to the selected music.
   - During the video, patients should mentally simulate themselves walking in sync with the rhythm (motor imagery).
   - Encourage attention to footstep timing relative to the bass drum and foot-ground contact relative to the bass line.
2. Internal Humming (Auditory-Motor Coupling)
   - Patients hum along internally with the melody while imagining their walking movements.

## Step 3: Walking Training (Active Gait Synchronization)

1. Assisted Phase (if needed)
   - For patients with gait freezing or instability, begin with stepping in place while holding onto a stable surface.
   - Gradually transition to assisted walking with therapist support.
2. Independent Walking
   - Patients walk naturally to the rhythm of the modified music.
   - Encourage them to match their footsteps to the bass drum beat for optimal rhythmic entrainment.
   - Monitor gait symmetry, step length, and fluidity of movement.

## Step 4: Post-Training Reflection & Adaptation

1. Self-Assessment

- Patients reflect on how well they could synchronize with the rhythm.
- Ask: *"Did the drum and bass components help guide your steps?" "Did you feel more stable when moving to the beat?"*

2. Therapist/Clinician Feedback
- Evaluate gait performance and make individualized adjustments for the next session.